\documentclass[article,reqno]{amsart}
\usepackage{amsmath,amssymb,mathtools}
\usepackage{latexsym}
\usepackage{epsfig}
\usepackage{booktabs}
\usepackage{color}
\usepackage{graphicx}
\usepackage{tikz}
\usepackage{array,booktabs}
\usepackage{pgfplots}
\usepackage{physics}
\newcommand{\beq}{\begin{equation}}
\newcommand{\eeq}{\end{equation}}
\newcommand{\beqnl}{\begin{eqnarray}}
\newcommand{\eeqnl}{\end{eqnarray}}

\newcommand{\be}{\begin{equation}}
\newcommand{\ee}{\end{equation}}
\newcommand{\bes}{\begin{equation*}}
\newcommand{\ees}{\end{equation*}}
\newcolumntype{C}{>{\centering\arraybackslash}m{1in}} 

\numberwithin{equation}{section}

\title{Analogue Hawking Effect: BEC and Surface Waves}
\author{F. Belgiorno$^{1,2,3}$, S.L. Cacciatori$^{2,4}$, A. Farahat$^{1}$, and A. Vigan\`o$^{2,5}$ }
\address{\noindent $^1$Dipartimento di Matematica, Politecnico di Milano, Piazza Leonardo 32, IT-20133 Milano, Italy\endgraf
$^2$INFN sezione di Milano, via Celoria 16, IT-20133 Milano, Italy\endgraf
$^3$INdAM-GNFM \endgraf
$^4$Department of Science and High Technology, Universit\`a dell'Insubria, Via Valleggio 11, IT-22100 Como, Italy\endgraf
$^5$Dipartimento di Fisica, Universit\`a degli Studi di Milano, Via Celoria 16, IT-20133 Milano, Italy\endgraf
}

\begin{document}

\maketitle

\begin{abstract}
We take into account two further physical models which play an utmost importance in the framework of Analogue Gravity. We first 
consider Bose--Einstein condensates (BEC) and then surface gravity waves in water.  Our approach is based on the 
use of the master equation we introduced in a previous work. A more complete analysis of the singular perturbation 
problem involved, with particular reference to the behavior in the neighbourhood of the (real) turning point 
and its connection with the WKB approximation, allows us to verify the thermal character of the particle production process. 
Furthermore, we can provide a simple scheme apt to 
calculate explicitly the greybody factors in the case of BEC and surface waves. 
This corroborates the improved approach we proposed for 
studying the analogue Hawking effect in the usual limit of small dispersive effects. 
\end{abstract}

\section{Introduction}

In~\cite{belmaster} a possible unifying formalism was proposed for dealing with the analogous 
Hawking effect, based on a fourth-order equation of the Orr--Sommerfeld type extensively studied 
in a series of papers by Nishimoto (see~\cite{ni,ni-I,ni-tp,ni-global}).

We take into consideration a further very interesting case, involving BEC and also surface waves,  
and provide both a sensible approximation scheme to the associated problems and also an analytical calculation of the greybody factor. 
We refer the reader to the following sections for an extensive list of references for theoretical studies of both the models. 
The utmost relevance of both the models is self-evident, due to the fact that BEC and water have been the most generally 
recognised  benchmarks for experimental verification of the actual existence of Hawking radiation {\color{black}~\cite{rousseaux-first,
weinfurtner-prl,rousseaux-book,weinfurtner-book,rousseaux,jeff-nature,denova,rousseaux-cocurrent}.}
We do not discuss herein the problem represented by the subcritical case, which is left for future investigations.

In the following, 
we first take into account the Hawking effect in BEC. We adopt the  healing length as the expansion parameter to be taken into account 
in order to deal with the problem of small dispersive effects. The well-known superluminal nature of the dispersion relation requires 
a different analysis with respect to the subluminal cases discussed in~\cite{belmaster}, with particular reference to the near horizon 
approximation. We obtain analytical solutions both in the far region and in the one near the turning point (horizon). 
We propose a different solution with respect to the ones existing in the literature, and obtain a complete analytical formula 
for inferring thermality and  the grey-body factor.

In the second part of the paper we consider gravity waves in water. As expansion parameter we consider the shallowness parameter 
and, in this subluminal case, we again perform, by following analogous theoretical paths as above, the calculation of thermality 
and a scheme for the calculation of the grey-body factor. 
A further discussion 
follows. 

\section{BEC}

There are several theoretical studies on analogous Hawking effect in BEC, starting from the seminal paper by Garay et al.~\cite{garay}. 
We limit ourselves to some relevant references concerning mainly semi-analytical/analytical approaches to the dispersive case~\cite{balbinot,macher-bec,larre,fabbri-hydro,mayoral-disp,anderson,balbinot-book,anderson-gray,anderson-low,anderson-exact,coutant-bdg}. 

We refer mainly to~\cite{macher-bec}. For completeness, we reproduce some basic steps towards the equations we study in the following subsections. The field $\hat{\Psi} (t, {\bf x})$ of atoms of the condensate satisfies the commutation relations
\beq
[\hat{\Psi} (t,{\bf x}),\hat{\Psi}^\dagger (t,{\bf x}')]=\delta^3  ({\bf x}-{\bf x}'),
\eeq
and the Heisenberg equation of motion
\beq
[\hat{\Psi} (t,{\bf x}),\hat{H}]=i\hbar \partial_t \hat{\Psi} (t,{\bf x}),
\eeq
where the Hamiltonian operator is 
\beq
\hat{H} =\int d^3 x \left[  \frac{\hbar^2}{2m} \nabla_{\bf x} \hat{\Psi}^\dagger \nabla_{\bf x} \hat{\Psi}+V \hat{\Psi}^\dagger \hat{\Psi}
+\frac{g}{2} \hat{\Psi}^\dagger \hat{\Psi}^\dagger  \hat{\Psi} \hat{\Psi}\right],
\eeq
where $m$ is the mass of the atoms, $V$ is the external potential, and $g$ is an effective coupling \cite{macher-bec}. 
At very low temperatures, a condensed state forms, represented by a (c-number) state $\Psi_0$, and one may introduce also a 
relative (quantum) fluctuation  in such a way that 
\beq
\hat{\Psi}=\Psi_0 (1+\hat{\phi}), 
\eeq
and then, considering only a 1-dimensional condensate henceforth, from the linearized equation one obtains 
\beq
\label{lin-eq}
i \hbar (\partial_t +v(x) \partial_x) \hat{\phi} =T \hat{\phi} + m c^2 (\hat{\phi}+\hat{\phi}^\dagger),
\eeq
where $v(x),c(x)$ are the local flow velocity and the speed of sound, respectively, and for stationary condensates 
\beq
T\coloneqq -\hbar^2 \frac{1}{2 m} v(x) \partial_x \biggl(\frac{1}{v(x)} \partial_x\biggr).
\eeq
Focusing only on stationary condensates, one gets
\beq
\hat{\phi}_\omega (t,x) = \hat{a}_\omega e^{-i\omega t} \phi_\omega (x)+\hat{a}_\omega^\dagger (e^{-i\omega t} \varphi_\omega (x))^\ast.
\eeq
Then from \eqref{lin-eq} and the commutation relations for $\hat{a}_\omega, \hat{a}_\omega^\dagger$ one obtains for the 
stationary modes $\phi_\omega$, $\varphi_\omega$, 
which will be indicated as $\phi$, $\varphi$ henceforth, satisfy
\begin{align}
\label{phi-varphi}
(\hbar (\omega+ i v(x) \partial_x) -T-m c^2 (x)) \phi (x) &= m c^2 (x) \varphi (x), \\
\label{varphi-phi}
(-\hbar (\omega+ i v(x) \partial_x) -T-m c^2 (x)) \varphi (x) &= m c^2 (x) \phi (x),
\end{align}
It is straightforward to show that one may obtain separated equations for $\phi$, $\varphi$, as in~\cite{macher-bec}: 
\beq
\label{phi-eq}
\biggl[(\hbar (\omega+ i v(x) \partial_x) +T) \frac{1}{c^2 (x)} (-\hbar (\omega+ i v(x) \partial_x) +T)+ 2 m T\biggr]
\phi (x) =0,
\eeq
and 
\beq
\label{varphi-eq}
\biggl[(-\hbar (\omega+ i v(x) \partial_x) +T) \frac{1}{c^2 (x)} (\hbar (\omega+ i v(x) \partial_x) +T)+ 2 m T\biggr]
\varphi (x) =0.
\eeq
Both the above equations are fourth order ones, and henceforth we first take into account~\eqref{phi-eq} for 
modes $\phi$.
We notice that we can simplify a factor $\hbar^2$ overall. Furthermore, in order to eliminate the 
third order term, we put
\beq
\label{third-phi}
\phi (x) = c(x) \sqrt{v(x)} \zeta (x). 
\eeq
Then we obtain the equation
\beq
\label{four}
\bigl[\alpha_4 (x) \partial_x^4 + \alpha_2 (x) \partial_x^2+ \alpha_1 (x) \partial_x +\alpha_0 (x)\bigr] \zeta (x)=0,
\eeq
where 
\begin{subequations}
\begin{align}
\alpha_4 (x) &= \frac{\hbar^2}{4 m^2 c^2 (x)},\\
\alpha_2 (x) &= -1+\frac{v^2 (x)}{c^2 (x)} + \ldots,\\
\alpha_1 (x) &= \frac{2}{c^2 (x)} \bigl( -i \omega v(x)- c(x) c'(x) +v(x) v'(x)\bigr)+\ldots\\   
\begin{split}
\alpha_0 (x) &= \frac{1}{c^2 (x)} \biggl(-\omega^2 -2 \frac{v^2 (x) (c'(x))^2}{c^2 (x)}-i \omega v'(x) +
\frac{v(x) v'(x) c'(x)}{c(x)}+\frac{(v'(x))^2}{4}+\frac{3 (v'(x))^2 c^2 (x)}{4 v^2 (x)} \\
& \quad -c''(x) c(x) +
\frac{v^2 (x) c''(x)}{c(x)}-\frac{v''(x) c^2 (x)}{2 v(x)}+\frac{v(x) v''(x)}{2}\biggr)+\ldots.
\end{split}
\end{align}
\end{subequations}
In the above formulas, we did not write explicitly all the terms. The complete expression of the coefficients 
appears in appendix~\ref{bec-coeffs}.

A natural expansion parameter suitable for a weakly dispersive regime is the so-called healing length
\beq
\xi\coloneqq\frac{\hbar}{\sqrt{2} m c (x)},
\eeq
which depends on the local speed of sound. Let us define 
\beq
\bar{\xi}=\sup_x \xi (x) = \frac{\hbar}{\sqrt{2} m} \frac{1}{\inf_x c (x)} \eqqcolon
\frac{\hbar}{\sqrt{2} m} \frac{1}{\bar{c}},
\eeq
where 
\beq
\bar{c}\coloneqq\inf_x c (x)>\epsilon>0.
\eeq
Of course we have $0<\xi (x)\leq\bar{\xi}$, and $\bar{\xi}\to 0$ has to be meant as the limit of 
weak dispersive effects we are interested in\footnote{It should be more correctly intended as the 
limit where the healing length is much smaller than the wavelength of the perturbation on the background condensate~\cite{balbinot}}.  
We obtain the following fourth order equation of the Orr--Sommerfeld type~\cite{ni}
\beq
\bigl[\bar{\xi}^2\partial_x^4 -\bigl( 
\beta_2 (x,\bar{\xi}) \partial_x^2+ \beta_1 (x,\bar{\xi}) \partial_x +\beta_0 (x,\bar{\xi})\bigr)\bigr] \zeta (x)=0,
\label{orso}
\eeq
where
\begin{subequations}
\begin{align}
\beta_2 (x,\bar{\xi}) &= \frac{2 c^2 (x)}{\bar{c}^2} \biggl(1-\frac{v^2 (x)}{c^2 (x)} \biggl) + 
\left( -\frac{i \sqrt{2} \bar{c}  v(x) c'(x)}{c^3(x)}+\frac{i \sqrt{2} \bar{c}  v'(x)}{c^2(x)}\right) \bar{\xi}+O(\bar{\xi}^2),\\
\beta_1 (x,\bar{\xi}) &= \frac{4}{\bar{c}^2} \left( i \omega v(x)+ c(x) c'(x) -v(x) v'(x)\right)+O(\bar{\xi}),\\
\begin{split}
\beta_0 (x,\bar{\xi}) &=\frac{2}{\bar{c}^2} \biggl(\omega^2 +2 \frac{v^2 (x) (c'(x))^2}{c^2 (x)}+i \omega v'(x) - 
\frac{v(x) v'(x) c'(x)}{c(x)}-\frac{(v'(x))^2}{4}-\frac{3 (v'(x))^2 c^2 (x)}{4 v^2 (x)} \\
& \quad + c''(x) c(x) -
\frac{v^2 (x) c''(x)}{c(x)}+\frac{v''(x) c^2 (x)}{2 v(x)}-\frac{v(x) v''(x)}{2}\biggr)+O(\bar{\xi}).
\end{split}
\end{align}
\end{subequations}
\subsection{The reduced equation}
The reduced equation is
\beq 
\bigl( 
\beta_2 (x,0) \partial_x^2+ \beta_1 (x,0) \partial_x +\beta_0 (x,0)\bigr) \zeta (x)=0,
\eeq
which displays a turning point (TP) such that 
\beq
\beta_2 (x_{TP},0)=0 \Longleftrightarrow \biggl(1-\frac{v^2 (x)}{c^2 (x)} \biggr)|_{x_{TP}}=0. 
\eeq
As in~\cite{macher-bec}, we can assume $x_{TP}=0$ and get a black hole horizon for
\beq
v(x)+c(x)=0,
\eeq
with $v<0$, and also in the linear region
\beq
\label{lin-region}
v(x)+c(x)\sim \kappa x,
\eeq
with {\color{black}$\kappa \coloneqq v'(0)+c'(0)>0$}.  
As to~\eqref{varphi-eq}, we point out that in the limit as $\bar{\xi}\to 0$, we obtain the 
same leading order contributions for $\varphi$ as for $\phi$. This is true for the results displayed 
in the following two subsections, so we shall not repeat the calculation also for $\varphi$.

\subsection{WKB approximation}
\label{becwkb}

We put 
\beq
\zeta (x) = \exp ( \frac{\theta(x)}{\bar{\xi}} ) \sum_{n=0}^\infty \bar{\xi}^n y_n (x).
\eeq
To the lowest order, we obtain 
\beq
{\theta'}^4 \bar{c}^2- 2(c^2 (x)-v^2 (x)) {\theta'}^2 =0, 
\eeq
whose solutions are $\theta'=0$ (multiplicity two), and  for $x<0$
\beq
\label{theta-phi}
\theta'_\pm = \pm i \frac{\sqrt{2}}{\bar{c}} \sqrt{v^2 (x)-c^2 (x)}.
\eeq
As expected, due to the superluminal nature of the dispersion relation in BEC, two big wavenumber 
modes are found in the black hole region $x<0$, where $v^2 (x)>c^2 (x)$.  We mention, in passing, 
that for $x>0$ the non-vanishing solutions correspond to the decaying mode and the growing mode respectively. As to the 
propagating solutions, we associate with them the so-called transport 
equation:
\beq
\label{transport-phi}
\begin{split}
& (v^2 (x)-c^2 (x)+\bar{c}^2 {\theta'}^2 (x)) y'_0 +\biggl(-i \omega v (x)- c(x) c'(x) +v(x) v'(x) \\
&+ i  \frac{\bar{c}}{\sqrt{2}} (v' (x)-v(x) \frac{c' (x)}{c(x)} ) \theta' (x)  -(c^2 (x)-v^2 (x)) \frac{\theta'' (x)}{2 \theta' (x)}
 +\frac{3}{2} \bar{c}^2 \theta'' (x) \theta' (x) \biggr) y_0 (x) =0.
\end{split}
\eeq
We then find the solutions 
\beq
y_{0 \pm} (x) = B_\pm \left( v^2 (x)-c^2 (x) \right)^{-3/4} 
\Biggl( \frac{v(x)}{c(x)}+\sqrt{\frac{v^2(x)}{c^2(x)}-1}\Biggr)^{\mp 1} 
\exp( - i \omega \int^x ds \frac{v (s)}{v^2 (s)-c^2 (s)}).
\eeq
In the near horizon region, it is easy to show that 
\beq
\label{wkb-lin}
|y_{0 \pm} (x)| \propto x^{-3/4},
\eeq
as usual and expected. The high momentum modes are then 
\beq
\phi_\pm (x) = c(x) \sqrt{v(x)} \zeta_\pm (x)= c(x) \sqrt{v(x)} y_{0 \pm} (x) \exp( \frac{\theta_\pm (x)}{\bar{\xi}}), 
\eeq
and $|\phi_\pm (x)| \propto x^{-3/4}$ near $x=0$, as in~\eqref{wkb-lin}. In particular, we have 
\beq
\label{wkb-linear}
\zeta_\pm^{wkb} (x) \sim - (2 c_0 \kappa)^{-3/4} |x|^{-\frac{i\omega}{2\kappa} -\frac{3}{4}} 
\exp( \mp \frac{2}{3} \frac{i}{\bar{\xi}}  \sqrt{\frac{4 c_0 \kappa}{\bar{c}^2}} |x|^{3/2} ),
\eeq
where $c_0 \coloneqq c(0)$; this formula will be useful in the following.

Two further solutions occurring when $\theta'=0$  can be obtained  from the reduced equation. First, we notice that near the turning point one obtains 
\beq
\biggl[\partial_x^2+ \frac{1}{x} \biggl(1-i \frac{\omega}{\kappa}\biggr) \partial_x +(\ldots)\frac{1}{x} \biggr] \zeta (x)=0, 
\label{phi-indicial}
\eeq
where the coefficient $(\ldots)$ does not contribute to the so-called indicial equation, 
whose roots are 
\begin{equation}
\alpha_1 =0, \quad
\alpha_2 = i \frac{\omega}{\kappa}.
\end{equation}
In particular, we can define~\cite{ni}
\beq
\lambda \coloneqq 1-\alpha_2 = 1-i \frac{\omega}{\kappa}.
\eeq 
We obtain near the regular singular point 
$x=0$ (our TP) the following series expansions:  for $x>0$ 
\begin{align}
\label{ser-wkb}
\phi_{v} (x) &= 1+\sum_{n=1}^{\infty} c_n x^n,\\
\phi_{u} (x) &= x^{i \frac{\omega}{\kappa}}  \biggl( 1+\sum_{n=1}^{\infty} d_n x^n\biggr) .
\end{align} 
The series expansion \eqref{ser-wkb} holds true on both sides of the turning points, with different coefficients. So we {\color{black} can obtain also 
analogous  expansions for the solutions $\phi_{d} (x)$ and $\phi_{l} (x)$ occurring for $x<0$. 
We omit the straightforward details.} 
By comparing the behavior of the above four solutions in the so-called linear region where~\eqref{lin-region} holds, with the solutions one can obtain in the near turning point approximation (to be discussed in the following subsection), one finds the connection 
formulas providing the amplitudes for pair-creation we are interested in. See the following.\\
 
In particular, it is useful to provide also approximate solutions of the reduced equation as $x$ is large (in the external region 
with respect to the black hole). It is easy to show that for large $x$ in the above sense we have 
$v(x)$, $c(x)\sim$const., and then $v'=0$, $c'=0$. The asymptotic values of $v(x)$, $c(x)$ as 
$x\to \infty$ are for simplicity indicated with $v_r$, $c_r$ respectively (analogously, one has $v\to v_l$, $c\to c_l$ for $x\to -\infty$).   As a consequence
e.g.~under the conditions of theorem 1.9.1 of~\cite{eastham}, we get asymptotically for $x\to \infty$
\begin{align}
\phi_{v} (x) &\sim \exp ( -i \frac{\omega}{c_r-v_r} x),\\
\phi_{u} (x) &\sim \exp ( i \frac{\omega}{c_r+v_r} x).
\end{align}
Analogously, for $x\to -\infty$ one obtains 
\begin{align}
\phi_{d} (x) &\sim \exp ( -i \frac{\omega}{c_l-v_l} x),\\
\phi_{l} (x) &\sim \exp ( i \frac{\omega}{c_l+v_l} x).
\end{align}
We note that $\phi_{l} (x)$ is a negative-norm mode.

\subsection{Approximation near the turning point}

Solutions near the TP have to satisfy the following equation, as shown in~\cite{belmaster}: 
\beq
\frac{d^4 \zeta}{dz^4}- \left( z \frac{d^2 \zeta}{dz^2} +\lambda \frac{d\zeta}{dz} \right)=0,
\label{four-z}
\eeq
where 
\beq
\label{lamb-super}
\lambda \coloneqq 1 - i \frac{\omega}{\kappa},
\eeq
and 
\beq
z=\left(\frac{4 c_0}{\bar{c}^2}\kappa\right)^{1/3} \epsilon^{-2/3} x, 
\eeq
where $c_0\coloneqq c(0)$.
There is a first solution which is constant, and put equal to one (cf.~\cite{ni}). This solution 
represents the near horizon approximation for the counter-propagating mode $v$ discussed in the 
previous subsection and, albeit nearly trivial, it is fundamental for getting a complete basis 
for solutions near the turning point.
Further solutions of equation~\eqref{four-z} can be found by means of Laplace integrals
\beq
\label{gen-airy}
\zeta_j(z) = \frac{1}{2\pi i} \int_{C_j} dt \; t^{\lambda -2} \exp (z t- \frac{1}{3} t^3),
\eeq
with a suitable choice for the paths $C_j$ in the complex $t$-plane. For the superluminal case at hand the aforementioned solutions of~\eqref{gen-airy} are also known as generalized Airy 
functions.

Paths extending to infinity in the complex $t$-plane must be restricted to allowed regions. We have the same regions as for the 
Airy functions, with $\theta\coloneqq\arg(t)$:
\beq
\theta\in \left(-\frac{\pi}{6},\frac{\pi}{6}\right) \cup \left(\frac{\pi}{2},\frac{5 \pi}{6}\right) \cup \left(\frac{7\pi}{6}, \frac{3\pi}{2}\right).
\eeq
The boundary conditions we introduce herein differ from the ones of the seminal investigation for 
a superluminal model contained 
in~\cite{corley}, and also from the ones in~\cite{unruh-s}. Some differences appear also with the 
analysis {\color{black} of~\cite{cpf,Coutant-thick} for the Corley model, where the same diagram was proposed, and} where, furthermore Fourier transform was used in place of Laplace transform. 

Our choice is the following: in order to describe the states in the external region $x>0$, which correspond to the decaying mode and to
the cut-mode related to the Hawking particle, we choose to introduce the cut in the positive 
real axis in the complex $t$-plane.

\begin{figure}[hbtp!]
\includegraphics[scale=0.55]{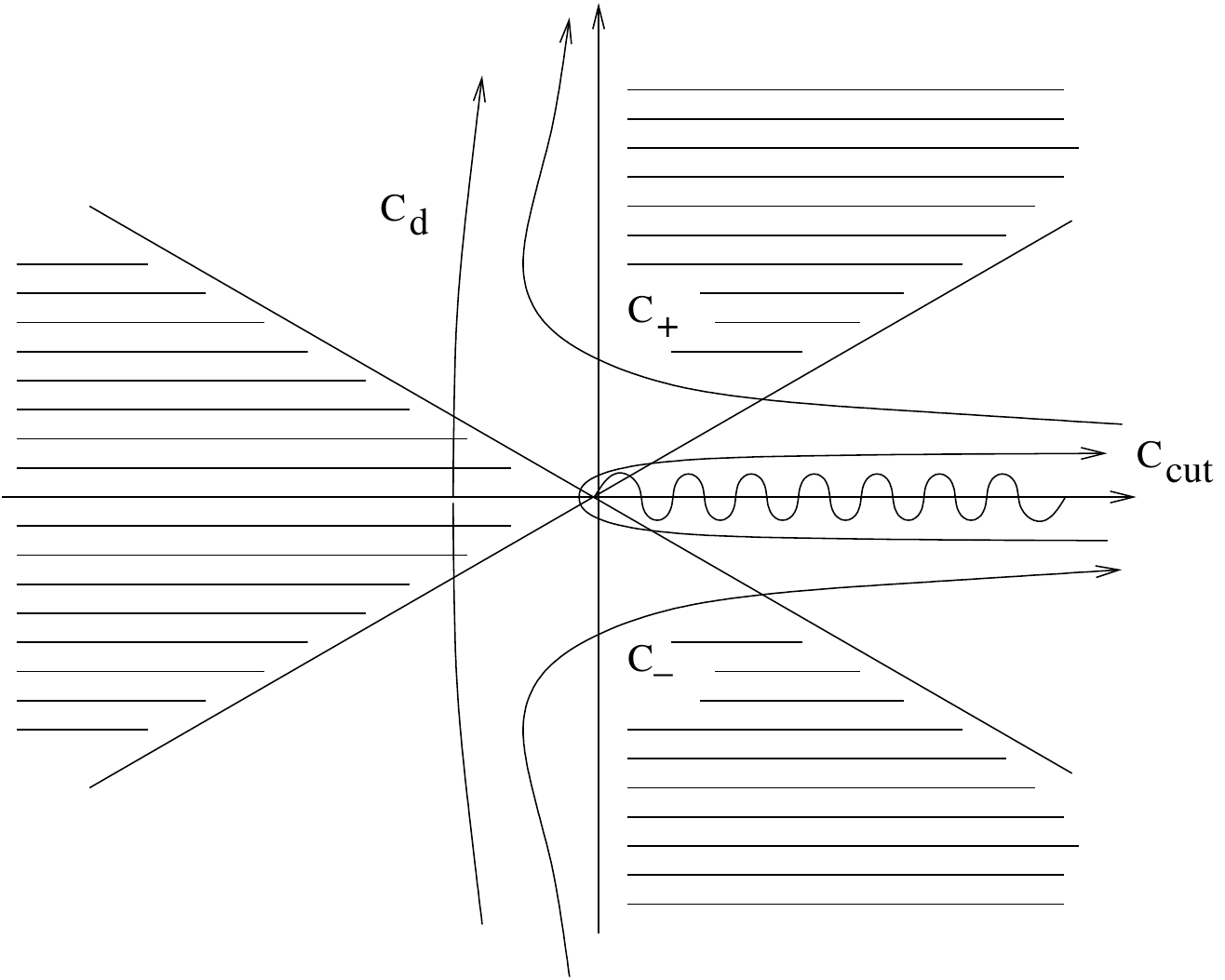} 
\caption{Paths for the superluminal case. $C_d$, $C_{cut}$ are the decaying mode and the Hawking mode respectively, 
and are found in the external region $x>0$. $C_\pm$ correspond to the dispersive modes, and are found in the 
black hole region $x<0$. 
$C_d$, $C_{cut}$ can be deformed into $C_+$, $C_-$.}
\label{fig:superl}
\end{figure}

One may consider as boundary condition the presence of the Hawking mode (cut mode) and of the decaying mode 
(albeit it does not participate to fluxes at infinity, 
it may play a role for local observables~\cite{cpf}). The corresponding paths are homotopic to the 
ones for the two states inside, which correspond to the big wavenumber $k_\pm$ states in the black hole region 
travelling towards the horizon. See also figure~\ref{fig:superl}.
We then obtain the `Corley's diagram' for the creation process of Hawking particles. In our interpretation,  at the level of the particle creation process, 
the fourth mode $v$ is not directly involved. 
Still, it plays a role in depleting the Hawking particle flux only in a further process of scattering on the geometry associated with the reduced equation, 
in analogy to our discussion for the cases of the subluminal Corley model and of the dielectric model taken into account  in~\cite{belmaster}. {\color{black} 
This is in agreement with the analysis in \cite{cpf,Coutant-thick} for the Corley model.} Cf.~\cite{belmaster}, figure 2 therein.

It is important to note that $\zeta_j (z)$ can be rewritten as follows: 
by putting $t=\sqrt{|z|} u$, we get 
\beq
\zeta_j(z) = \frac{1}{2\pi i} |z|^{\frac{\lambda-1}{2}} I_j (z),
\eeq
where 
\beq
I_j (z) = \int_{\bar{C}_j} du \; g(u) \exp (|z|^{3/2} h_\pm (u)),
\eeq
and
\begin{align}
g(u)&\coloneqq u^{\lambda-2},\\
h_\pm (u)&\coloneqq \pm u-\frac{u^3}{3}.
\end{align}
For the decaying mode, which passes through the saddle point at $u=1$ (in the external region 
$h_+$ is involved) we get
\beq
\label{decay-super}
\zeta_{decaying} (z)\simeq e^{-i \pi} e^{\frac{\pi \omega}{\kappa}} \frac{1}{2 \sqrt{\pi}} |z|^{-\frac{i\omega}{2\kappa}-\frac{3}{4}} e^{-\frac{2}{3} |z|^{3/2}}. 
\eeq
As to the cut mode, we stress that the branch cut lies along a steepest descent. Indeed, we have that the imaginary part of $u-u^3/3$ is 
$b(1-a^2+b^2/3)$, where $a,b$ are the real part and the imaginary part of $u$ respectively. As a consequence,
$b=0$ is a steepest descent line. This allows us to calculate the cut contribution along the lines suggested in~\cite{miller}, chapter 4, section 4.8, finding thus
\beq
\label{cut-super}
\zeta_{cut} (z) \simeq -\frac{1}{i \pi} |z|^{i \frac{\omega}{k}} \Gamma \Bigl(-i\frac{\omega}{\kappa}\Bigr) \sinh (\frac{\pi \omega}{\kappa}). 
\eeq
For $x<0$ we have the modes $k_\pm$ in correspondence of the steepest descents passing through the saddle points 
$u_\pm = \pm i$.  
Then we get {\color{black} (cf. also \cite{corley,cpf,Coutant-thick} for the Corley model), with a coincidence that is related to the universal character 
of our near-horizon equation, which holds true also for the Corley model \cite{belmaster} }
\begin{align}
\label{saddle-m-super}
\zeta_- (z) &\simeq \frac{1}{2 \sqrt{\pi}} e^{\frac{1}{4}\pi i} e^{-\frac{\pi \omega}{2 \kappa}} |z|^{-\frac{i\omega}{2\kappa}-\frac{3}{4}} e^{i\frac{2}{3} |z|^{3/2}}, \\
\label{saddle-p-super}
\zeta_+ (z) &\simeq -\frac{1}{2 \sqrt{\pi}} e^{\frac{3}{4}\pi i} e^{\frac{\pi \omega}{2 \kappa}} |z|^{-\frac{i\omega}{2\kappa}-\frac{3}{4}} e^{-i\frac{2}{3} |z|^{3/2}}.
\end{align}
The black hole boundary condition near the horizon forces the coefficients of the 
modes $\pm$ and $u$ to be equal; by appealing to the Cauchy theorem, and to the fact that all the functions in the near-horizon 
approximation are well-defined for $z=0$, we have  
\beq
\zeta_+ (0) + \zeta_- (0) +\zeta_u (0)=\zeta_{decaying} (0).  
\eeq
As discussed in \cite{belmaster}, this condition amounts to the Corley's ansatz, i.e. the black hole boundary condition. 
 The fourth mode does not appear in the diagram, and its {\color{black} contribution is zero at the level of the pair-creation process}. 
 See \cite{belmaster} for a further discussion. 
 See also below.
Then, by comparing with the WKB solutions in the matching region, we find for the propagating part of the field
\beq
\label{connection}
\begin{split}
\phi (x,t) &= \bigg(e^{\frac{3}{4}\pi i} \frac{e^{{\frac{\pi \omega}{2\kappa}}}}{2\sqrt{\pi}} 
2^{1/4} \sqrt{c_0 \kappa \bar{c}} \left(\frac{4 c_0 \kappa}{\bar{c}^2}\right)^{-\frac{i\omega}{6\kappa}} 
\bar{\xi}^{\frac{i\omega}{3\kappa}+\frac{1}{2}}\phi_+ (x,t) \\
& \quad +
e^{\frac{1}{4}\pi i} \frac{e^{-{\frac{\pi \omega}{2\kappa}}}}{2\sqrt{\pi}} 
2^{1/4} \sqrt{c_0 \kappa \bar{c}} 
\left(\frac{4 c_0\kappa}{\bar{c}^2}\right)^{-\frac{i\omega}{6\kappa}} 
\bar{\xi}^{\frac{i\omega}{3\kappa}+\frac{1}{2}}\phi_- (x,t) \bigg) \theta (-x)\\
& \quad 
+\bigg(-\frac{\sinh (\frac{\pi \omega}{ \kappa}) }{\pi i} \Gamma \Bigl(- i \frac{\omega}{ \kappa}\Bigr)
\left(\frac{4 c_0\kappa}{\bar{c}^2}\right)^{\frac{i\omega}{3\kappa}} \bar{\xi}^{-\frac{2i\omega}{3\kappa}}
\phi_{u} (x,t)+ h \phi_{v} (x,t)\bigg) \theta (x), 
\end{split}
\eeq
where $\phi_\pm$, $\phi_u$, $\phi_v$ are the WKB solutions for the modes at hand. % and the $h$ is {\color{black} actually zero}. 
As to the modes $d,l$, one may proceed as in the Corley model discussed in \cite{belmaster}. We do not delve 
into the details. 

\subsection{Calculations for the $\varphi$ stationary modes}

In place of~\eqref{phi-eq} one must consider~\eqref{varphi-eq}, and again the 
parameter for asymptotic expansion is $\bar{\xi}$. We do not repeat all the steps, 
and we limit ourselves to point out some features. In order to eliminate the third order term, 
one puts again 
\beq
\label{third-varphi}
\varphi (x) = c(x) \sqrt{v(x)} \eta (x).
\eeq
As to the WKB approximation, we note that equation for $\theta$ remains the same as for $\phi$, and that for the 
long wavenumber modes one obtains again~\eqref{theta-phi}.
For the transport equation only a change occurs, 
\beq
\label{transport-varphi}
\begin{split}
& (v^2 (x)-c^2 (x)+\bar{c}^2 {\theta'}^2 (x)) y'_0 +\biggl(-i \omega v (x)- c(x) c'(x) +v(x) v'(x) \\
&- i  \frac{\bar{c}}{\sqrt{2}} (v' (x)-v(x) \frac{c' (x)}{c(x)} ) \theta' (x)  -(c^2 (x)-v^2 (x)) \frac{\theta'' (x)}{2 \theta' (x)}
 +\frac{3}{2} \bar{c}^2 \theta'' (x) \theta' (x) \biggr) y_0 (x) =0.
\end{split}
\eeq
We then find the solutions 
\beq
y_{0 \pm} (x) = B \left( v^2 (x)-c^2 (x) \right)^{-3/4} 
\left( \frac{v(x)}{c(x)}+\sqrt{\frac{v^2(x)}{c^2(x)}-1}\right)^{\pm 1} 
\exp ( - i \omega \int^x ds \frac{v (s)}{v^2 (s)-c^2 (s)}).
\eeq
This does not substantially modify the expansion in the linear region. As to the reduced equation, one has again 
\beq
\label{varphi-indicial}
\biggl[\partial_x^2+ \frac{1}{x}\biggl(1-i \frac{\omega}{\kappa}\biggr) \partial_x +\frac{1}{x}(\ldots)\biggr] \eta (x)=0, 
\eeq
whose indicial equation is the same as in the previous sections, and also the solutions remain the 
same in the asymptotic region.

Solutions near the TP have to satisfy the same equation as for the other mode {\color{black} \eqref{four-z}, simply through the 
substitution $\zeta\mapsto \eta$.}  As a consequence, also the near horizon 
solutions remain the same, and also the matching formulas in the linear region do not change. 

\subsection{Thermality}
 
We recall that with the stationary modes $\phi$, $\varphi$ one can associate conserved currents (see~\cite{dalfovo} and~\cite{larre} for an application to the analogous Hawking radiation):
\beq
J_x^{[\phi_a,\varphi_a]} \coloneqq - i \frac{\hbar}{2 m} \left( \phi_a^\ast \partial_x \phi_a- 
\phi_a \partial_x \phi_a^\ast+\varphi_a^\ast \partial_x \varphi_a- 
\varphi_a \partial_x \varphi_a^\ast \right),
\eeq
where we have $a=\pm,u,v$. In the following, we indicate simply with $J_x$ the above currents, and 
we mean to exploit the following current flux conservation: 
\beq
\label{fluxes}
|J_x^u|=|J_x^+|-|J_x^-|+|J_x^v|,
\eeq
where the outgoing flux of Hawking particles ($u$-modes, directed towards $\infty$) originates from the 
ingoing flux of modes ($v$ and $k_\pm$ modes, directed towards the horizon $x=0$), and the nature of the modes $k_-$ to be 
negative norm modes has be taken into account.

The normalization 
to the modes  is as in~\cite{macher-bec,balbinot-book}, by requiring that in the eikonal approximation the dispersion relation (which holds with constant coefficients in the asymptotic regions) 
\beq
\label{dispersion}
(\omega - v k)^2 = c^2 k^2 \left(1+ \frac{\xi^2}{2} k^2\right) ,
\eeq
holds true. The WKB solutions, as $|x|\to \infty$, behave as plane waves which we indicate as $\phi_\omega$, $\varphi_\omega$ for simplicity (omitting for the moment any further mode label). We take into account that 
the two components $(\phi_\omega,\varphi_\omega)$ satisfy the equations of motion~\eqref{phi-varphi} and~\eqref{varphi-phi}, 
and then we get~\cite{balbinot-book}
\begin{align}
\phi_\omega  &= D_\omega e^{-i \omega t+i k(\omega) x}= N_\omega \biggl(\omega -v k + c \frac{\xi}{\sqrt{2}} k^2\biggr) e^{-i \omega t+i k(\omega) x},\\
\varphi_\omega  &= E_\omega e^{-i \omega t+i k(\omega) x}=N_\omega \biggl(-\biggl(\omega -v k - c \frac{\xi}{\sqrt{2}} k^2\biggr) \biggr) e^{-i \omega t+i k(\omega) x},
\end{align}
where
\beq
N_\omega = \frac{1}{\sqrt{4 \sqrt{2} \pi \hbar \rho c \xi k^2 \abs{(\omega -v k) \left( \frac{d k(\omega)}{d\omega} \right)^{-1}}}},
\eeq
with $\rho \propto 1/v$. 
These normalization factors in the asymptotic region reduce to the ones of the homogeneous BEC, of course.

For explicit calculations, we point out that for each mode it holds 
\beq
\abs{J_x} \propto k (|D_\omega|^2+ |E_\omega|^2) \propto (\omega-v k)^2+ \left(c \frac{\xi}{\sqrt{2}} k^2 \right)^2.
\eeq
As usual, for thermality {\color{black} $\frac{|J_x^{-}|}{|J_x^{+}|} = e^{-\beta \omega}$ holds, where $\beta= \frac{2 \pi}{\kappa}$ is the inverse Hawking temperature}.

\subsection{Grey-body factor}

As to the grey-body factor, in \cite{belmaster} it has also been shown that, in principle, one might deduce the grey-body factor from the direct calculation of 
\beq
|\beta_\omega|^2 \coloneqq \frac{|J_x^{-}|}{|J_x^{u}|},
\eeq
which represents the number of created particles, as well-known. Even if this route is viable, the drawback is that there is the risk of a poor approximation 
(as in the standard Hawking effect calculations).\\ 
The grey-body factor can be obtained as  follows:
\beq
\Gamma = 1 -R, % \frac{|J_x^{v}|}{|J_x^{u}|},
\eeq
where we also define the ratio
\beq
R \coloneqq \frac{|J_x^{v}|}{|J_x^{u}|}
\eeq
As discussed in \cite{belmaster}, actually the mode $v$ does not participate {\sl directly} to the Hawking pair-creation {\color{black} process. Still, there can be a further contribution to $R$ arising from the back-scattering on the geometry,  %such that $R=R_{geom}$, 
leaving room for $\Gamma<1$. Then, $R$ represents  the reflection coefficient for the scattering of Hawking particles in the background geometry associated with the reduced equation obtained for $\bar{\xi}=0$.  This is in agreement with what happens in the Corley model \cite{cpf}.}
Given a $u$-mode entering from the 
part of the linear region, where the WKB approximation is valid, the reduced equation provides the contribution 
\beq
R_{reduced} \coloneqq  \left(\frac{|J_x^{v}|}{|J_x^{u}|}\right)_{reduced},
\eeq
with the fluxes computed asymptotically, and with $|J_x^{v}|$ measured 
near the horizon, but still in a region where the WKB works well). See \cite{belmaster} for a complete discussion. 
As the aforementioned geometry  amounts to the 
classical geometry for BEC analogous black holes, we refer to the expressions already 
present in the literature and calculated in the so-called hydrodynamic limit, see in particular~\cite{anderson-exact}. 
Of course, also in this case  there exists a maximal frequency $\omega_{max}$ such that, for $\omega > \omega_{max}$, 
only two modes participates to the scattering process and the Hawking effect is no more present~\cite{macher-bec}, 
so that  the spectrum is truncated at $\omega_{max}$ for non-zero values of $\bar{\xi}$.

\section{Shallow water waves}

Shallow water waves are the other fundamental benchmark of analogue gravity as, just for the 
case of BEC, experimental measurements of the analogous Hawking effect were carried out~\cite{weinfurtner-prl,weinfurtner-book,rousseaux}. Theoretical studies start with~\cite{schutzhold} and have been deepened further on. 
Also the phenomenon of undulation has been studied 
in detail~\cite{coutant-parentani-fluid}, as well as the problem of the subcritical case~\cite{euve,parentani-sub,coutant-sub,coutant-kdv}. We discuss herein only the transcritical case, which is the 
one properly associated with the analogous Hawking effect (although the subcritical case may preserve some 
imprinting of the Hawking phenomenon~\cite{coutant-sub}). Furthermore, we refer to the model discussed in 
~\cite{coutant-sub}.

As in~\cite{coutant-sub} (cf. also~\cite{coutant-parentani-fluid}), we limit ourselves to the weakly dispersive case where 
\beq
\tanh (i h \partial_x ) \mapsto i h\partial_x +i\frac{1}{3} h^3 \partial_x^3.
\eeq
$h$ is the local height (depth) of water. 
The corresponding (approximate) action is \cite{coutant-sub}
\beq
S=\frac{1}{2} \int d^2x [((\partial_t +v(x) \partial_x) \phi)^2 -c^2(x) (\partial_x \phi)^2+\frac{g h^3(x)}{3} (\partial_x^2 \phi)^2],
\eeq
where $v(x)$ is the local velocity of the fluid, $c(x)$ is the local speed of sound, $h(x)$ is the 
local height and $g$ is the gravity acceleration. 
This is the case most similar to the original model studied in~\cite{corley,cpf}, and is characterized by a subluminal 
dispersion, as well known.  
The equation of motion for stationary modes $\phi (t,x)=e^{-i\omega t} \psi (x)$ is a quartic equation of the following form: 
\beq
\label{eq-sw}
\left(\frac{g}{3} h^3 \partial_x^4+ 2 g h^2 h' \partial_x^3 + [ (c^2-v^2) +g (2 h (h')^2+h^2 h'')]
\partial_x^2+2 (i\omega v+c c' -v v') \partial_x +(\omega^2 + i v' \omega) \right)\psi=0.
\eeq
It is also to be noted that the model is associated with a 
peculiar conserved current for stationary modes
\beq
\label{curr-sw} 
J_x = \Im \biggl[i \omega v \psi^\ast \psi + (c^2-v^2) \psi^\ast \partial_x \psi+\frac{g}{3} \psi^\ast 
  \partial_x (h^3 \partial_x^2 \psi) -\frac{g}{3} h^3 (\partial_x \psi^\ast) (\partial_x^2 \psi)\biggr], 
\eeq
as in~\cite{coutant-sub,prain}.

\subsection{Rescaled variables}

It is useful to proceed as in~\cite{johnson-book}, by defining rescaled adimensional variables $z,\tau$ in place of $x,t$ as follows:
\begin{align}
x &= \lambda s,\\
t &= \frac{\lambda}{\sqrt{g h_0}} \tau,
\end{align}
where $\lambda$ stays for the wavelength and $h_0$ is to be considered a reference height (we could assume, for 
example, $h_0 =\inf_x h(x)>0$. See also below). We can also introduce 
\begin{align}
v (x) &= \sqrt{g h_0}\; \bar{v}(x),\\
c (x) &= \sqrt{g h_0}\; \bar{c}(x),\\
\omega &= \frac{\sqrt{g h_0}}{\lambda}\; \bar{\omega},
\end{align}
as well as the so-called long wavelength or shallowness parameter~\cite{johnson-book}
\beq
\delta\coloneqq\frac{h_0}{\lambda}. 
\eeq
As a consequence, from equation~\eqref{eq-sw}, we obtain 
\beq
\label{adi-sw}
\begin{split}
& \biggl( \delta^2 (\partial_{s}^4+ 6\frac{h'}{h} \partial_s^3)+ \\
& \left[ 3 \frac{h_0^3}{h^3} (\bar{c}^2-\bar{v}^2) +\delta^2 \left( 6 \frac{(h')^2}{h^2}+3\frac{h''}{h}\right)\right]
\partial_{s}^2+6 \frac{h_0^3}{h^3} (i \bar{v} \bar{\omega} +\bar{c} \bar{c}' -\bar{v} \bar{v}') 
\partial_{s} +3 \frac{h_0^3}{h^3} (\bar{\omega}^2 + i \bar{v}' \bar{\omega}) \biggr)\psi=0,
\end{split}
\eeq
where, with some liberal attitude, a prime indicates the derivative with respect to the dimensionless variable $s$.
The third-order term can be removed by means of the following Liouville-like transformation
\beq
\psi = h^{-3/2} \zeta,
\eeq
which allows to obtain 
\beq
\label{orso-sw}
\bigl[\delta^2\partial_s^4 + 
\gamma_2 (s,\delta) \partial_s^2+ \gamma_1 (s,\delta) \partial_s +\gamma_0 (s,\delta)\bigr] \zeta (s)=0,
\eeq
where
\begin{subequations}
\begin{align}
\gamma_2 (s,\delta) &= 3 \frac{h_0^3}{h^3 (s)} \left(\bar{c}^2 (s)-\bar{v}^2 (s)\right) + O(\delta^2),\\
\gamma_1 (s,\delta) &= 6 \frac{h_0^3}{h^3 (s)} (i \bar{v}(s)\bar{\omega} +\bar{c}(s) \bar{c}' (s)-\bar{v}(s) \bar{v}'(s))-9  \frac{h_0^3 h'(s)}{h^4 (s)} \left(\bar{c}^2 (s)-\bar{v}^2 (s)\right)  + O(\delta^2),\\
\begin{split}
\gamma_0 (s,\delta) &= 3 \frac{h_0^3}{h^3 (s)} (\bar{\omega}^2 + i \bar{v}'(s) \bar{\omega}) 
+9 \frac{h_0^3}{h^4 (s)}\biggl[ -i  \bar{\omega} \bar{v}(s) h'(s)- (\bar{c}(s) \bar{c}' (s)- \bar{v}(s) \bar{v}' (s)) h'(s) \\
& \quad -\frac{1}{2} \left(\bar{c}^2 (s)-\bar{v}^2 (s)\right) \left(h''(s) +\frac{5}{2} \frac{(h'(s))^2}{h^2(s)}\right) \biggr]
+O(\delta^2).
\end{split}
\end{align}
\end{subequations}
\subsection{The reduced equation} The reduced equation is
\beq
\label{redux}
\left( 
\gamma_2 (s,0) \partial_s^2+ \gamma_1 (s,0) \partial_s +\gamma_0 (s,0)\right) \zeta (s)=0,
\eeq
which displays a turning point (TP) such that
\beq
\gamma_2 (s_{TP},0)=0 \Longleftrightarrow \left(\bar{v}^2 (s)-\bar{c}^2 (s) \right)|_{s_{TP}}=0. 
\eeq
As usual, by returning momentarily to dimensionful variables, we can assume $x_{TP}=0$ and get a black hole horizon 
for $v(x)+c(x)=0$,nwith $v<0$, and also in the linear region $v(x)+c(x)\sim \kappa x$, with $\kappa = v'(0)+c'(0)$. 
Also in this case near the TP one obtains 
\beq
\biggl[\partial_x^2+ \frac{1}{x} \biggl(1-i \frac{\omega}{\kappa}\biggr) \partial_x +\frac{1}{x}  (\ldots)\biggr] \zeta (x)=0, 
\eeq
where the coefficient $(\ldots)$ does not contribute to the so-called indicial equation, 
whose roots are again
\begin{equation}
\alpha_1 =0, \quad
\alpha_2 = i \frac{\omega}{\kappa}.
\end{equation}
and, again
\beq
\label{mu-ni}
\mu \coloneqq 1-\alpha_2 = 1-i \frac{\omega}{\kappa}.
\eeq 

\subsection{WKB approximation} 

We put 
\beq
\zeta (s) = \exp ( \frac{\theta(s)}{\delta} ) \sum_{n=0}^\infty \delta^n y_n (s).
\eeq
To the lowest order, we obtain 
\beq
{\theta'}^4 + 3 \left(\frac{h_0}{h}\right)^3 (\bar{c}^2 -\bar{v}^2) {\theta'}^2 =0, 
\eeq
whose solutions are $\theta'=0$ (multiplicity two), and 
\beq
\theta'_\pm = \pm i \sqrt{3\left(\frac{h_0}{h}\right)^3}\sqrt{\bar{c}^2 -\bar{v}^2}.
\label{theta-psi}
\eeq
We first take into account the latter solutions, and associate with them the so-called transport 
equation
\beq
\label{transport-psi}
\biggl(6 \biggl(\frac{h_0}{h}\biggr)^3 (\bar{c}^2 -\bar{v}^2)+ 4  {\theta'}^2\biggr) y'_0+\biggl(6 {\theta'} {\theta''}
 + 3 \biggl(\frac{h_0}{h}\biggr)^3
\biggl( 2 i \bar{v} \bar{\omega}+2 (\bar{c} \bar{c}'-\bar{v} \bar{v}') +\biggl(\frac{\theta''}{\theta'}- 3\frac{h'}{h}\biggr) (\bar{c}^2 -\bar{v}^2)\biggr) \biggr)y_0=0.
\eeq
We then find the solutions 
\beq
\label{long}
y_{0 \pm} (s) = B \left( \bar{c}^2 -\bar{v}^2 \right)^{-3/4} h^{9/4} \exp (i\bar{\omega} \int^s \frac{\bar{v}}{\bar{c}^2 -\bar{v}^2 }).
\eeq
As to the degenerate solutions with $\theta'=0$, as known, they must solve the reduced equation with $\delta=0$~\eqref{redux}.
We obtain near the regular singular point 
$s=0$ (our TP) the series expansions for $s>0$ 
\begin{align}
\psi_{v} (s) &= 1+\sum_{n=1}^{\infty} c_n s^n,\\
\psi_{u} (s) &= s^{i \frac{\omega}{\kappa}}  \biggl( 1+\sum_{n=1}^{\infty} d_n s^n\biggr) .
\end{align}
In particular, it is useful to provide also approximate solutions of the reduced equation as $s$ is large (in the external region 
with respect to the black hole). It is easy to show that for large $s\to \infty$ in the above sense, we have 
$v(x)$, $c(x)\sim$const., and then $v'=0$, $c'=0$. The asymptotic values of $\bar{v}(x)$, $\bar{c}(x)$ as 
$x\to \infty$ are for simplicity indicated with $\bar{v}$, $\bar{c}$ respectively. As a consequence
e.g.~under the conditions of theorem 1.9.1 of~\cite{eastham}, we get asymptotically for $s\to \infty$
\begin{align}
\psi_{v} (s) &\sim \exp ( -i \frac{\bar{\omega}}{\bar{c}-\bar{v}} s),\\
\psi_{u} (s) &\sim \exp ( i \frac{\bar{\omega}}{\bar{c}+\bar{v}} s).
\end{align}
As in the previous model, we can also study solutions for $s<0$, and the only propagating ones asymptotically behave as 
\begin{align}
\psi_{d} (s) &\sim \exp ( -i \frac{\bar{\omega}}{\bar{c}_l-\bar{v}_l} s),\\
\psi_{l} (s) &\sim \exp ( i \frac{\bar{\omega}}{\bar{c}_l+\bar{v}_l} s), 
\end{align}
where $\bar{v}_l$, $\bar{c}_l$ are the limits of $\bar{v}(s)$, $\bar{c}(s)$ for  
$x\to -\infty$ respectively.

\subsection{Near-horizon approximation}

Near the TP the following equation holds true (cf.~\cite{belmaster}): 
\beq
\label{four-z-sub}
\frac{d^4 \zeta}{dz^4}+ \left( z \frac{d^2 \zeta}{dz^2} +\mu \frac{d\zeta}{dz} \right)=0,
\eeq
where $\mu$ is given in~\eqref{mu-ni},
and 
\beq
z= \frac{h_0}{h(0)} \left(6 \bar{\kappa} \bar{c}_0 \right)^{1/3} \delta^{-2/3} s,
\eeq
where $\bar{\kappa}\coloneqq \kappa \sqrt{g h_0}/\lambda$ and $\bar{c}_0\coloneqq\bar{c}(0)$. We may also choose $h_0=h(0)$, i.e. we can assume that 
$h_0$ is the value of $h$ at the TP.

Apart for the constant solution, which is again put equal to one (cf.~\cite{ni}), 
further solutions of equation~\eqref{four-z-sub} can be found by means of Laplace integrals as in~\cite{corley,unruh-s}
\beq
\zeta_j(z) = \frac{1}{2\pi i} \int_{C_j} dt \; t^{\mu -2} \exp (z t+\frac{1}{3} t^3),
\label{gen-sub}
\eeq
with a suitable choice for the paths $C_j$ in the complex $t$-plane. In this case, a subluminal character of the nonlinear part 
is present.
Paths extending to infinity in the complex $t$-plane must be restricted to allowed regions, displayed in~\cite{corley,belmaster} 
and, with $\theta\coloneqq\arg(t)$, we obtain:
\beq
\theta\in \left(\frac{\pi}{6},\frac{\pi}{2}\right) \cup \left(\frac{5\pi}{6},\frac{7 \pi}{6}\right) \cup \left(\frac{3\pi}{2}, \frac{11\pi}{6}\right).
\eeq
We start from~\eqref{gen-sub} and follow the general method explained in~\cite{belmaster}. 
{\color{black} 
Paths can be chosen as in any subluminal case, cf. figure 2 in~\cite{belmaster}.  
It corresponds to the so-called black hole boundary condition as discussed in~\cite{corley}. See also \cite{cpf,Coutant-thick}. 
We limit ourselves to reproduce the results. We have for $x<0$ the decaying mode %passing through the saddle point $u=1$
\beq
w_{decaying} (z)\simeq \frac{1}{2 \sqrt{\pi}} |z|^{-\frac{i\omega}{2\kappa}-\frac{3}{4}} e^{-\frac{2}{3} |z|^{3/2}}. 
\label{decay-sub}
\eeq
As usual~\cite{corley}, it provides the aforementioned black hole boundary condition. 
For $x>0$ we have the modes $k_\pm$ in correspondence of the steepest descents}
%passing through the saddle points $u_\pm = \pm i$:
\begin{align}
\label{saddle-p-sub}
w_+ (z) &\simeq \frac{1}{2 \sqrt{\pi}} e^{-\frac{3}{4}\pi i} e^{\frac{\pi \omega}{2 \kappa}} |z|^{-\frac{i\omega}{2\kappa}-\frac{3}{4}} e^{i\frac{2}{3} |z|^{3/2}}, \\
\label{saddle-m-sub}
w_- (z) &\simeq \frac{1}{2 \sqrt{\pi}} e^{\frac{1}{4}\pi i} e^{-\frac{\pi \omega}{2 \kappa}} |z|^{-\frac{i\omega}{2\kappa}-\frac{3}{4}} e^{-i\frac{2}{3} |z|^{3/2}}.
\end{align}
The cut contribution, on the negative real axis, represents the Hawking mode and can be also in this case  calculated  along the lines suggested in~\cite{miller}, chapter 4, section 4.8:
\beq
w_{cut} (z) \simeq -\frac{1}{i \pi} |z|^{i \frac{\omega}{k}} \Gamma \Bigl(-i\frac{\omega}{\kappa}\Bigr) \sinh (\frac{\pi \omega}{\kappa}).
\label{cut-sub}
\eeq
We are interested in connecting the expansions of the near horizon approximation with the ones of the WKB approximation displayed 
in the previous section. We get 
\beq
\psi (s,\tau) = c_+ \psi_+ (s,\tau) + c_- \psi_- (s,\tau) + c_u \phi_u (s,\tau)+c_v \phi_v (s,\tau), 
\label{field}
\eeq
where {\color{black} the fourth mode gives no direct contribution to the pair-creation associated with the Hawking effect. Cf. \cite{belmaster}.}  We also have  
\beq
\psi_j (s,\tau) = \exp \left( -i \bar{\omega} \tau + i \bar{k}_j (\omega) \tau \right),
\eeq
with $j=\pm u,v$ and $\bar{k}_j (\omega)$ is the corresponding rescaled wavenumber. 
By comparing with the WKB solutions 
again in the matching region 
\beq
\label{connection-surf}
\begin{split}
\psi (s,\tau) &=
e^{-\frac{3}{4}\pi i}  (h(0))^{-9/4} 3^{-1/4} \sqrt{2 \bar{c_0} \bar{\kappa}} \frac{e^{\frac{\pi \omega}{\kappa} }}{2 \sqrt{\pi}} (6\bar{c_0} \bar{\kappa})^{-\frac{i \omega}{6  \kappa}}
\delta^{\frac{i\omega}{3  \kappa}+\frac{1}{2}}\psi_+ (s,\tau) \\ 
& \quad +e^{\frac{1}{4}\pi i}  (h(0))^{-9/4} 3^{-1/4} \sqrt{2 \bar{c_0} \bar{\kappa}}\frac{e^{-\frac{\pi \omega}{\kappa} }}{2 \sqrt{\pi}} (6\bar{c_0} \bar{\kappa})^{-\frac{i \omega}{6  \kappa}}
\delta^{\frac{i\omega}{3  \kappa}+\frac{1}{2}}\psi_- (s,\tau) \\
& \quad
-\frac{\sinh (\frac{\pi \omega}{\kappa}) }{\pi i} \Gamma \Bigl(-  \frac{i\omega}{\kappa}\Bigr) 
(6\bar{c_0} \bar{\kappa})^{\frac{i\omega}{3  \kappa}} 
\delta^{-\frac{2i\omega}{3  \kappa}}
\psi_{u} (s,\tau)+ c_v \psi_{v} (s,\tau).
\end{split}
\eeq
For $\psi_{d},\psi_{l}$ the matching is analogous. {\color{black} Only the scattering on the geometry may contribute to $c_v$. See below.}

\subsection{Thermality and grey-body factor} 

We note that the conserved current (\ref{curr-sw}) can be rewritten in terms of rescaled variables as follows: 
\beq
J_s \propto \Im \left(
i \bar{\omega} \bar{v}  \psi^\ast \psi + 
(\bar{c}^2-\bar{v}^2) \psi^\ast \partial_s \psi+
\frac{1}{3} \delta^2 \left( \psi^\ast \partial_s \left(
\left(\frac{h}{h_0}\right)^3 \partial_s^2 \psi \right) - \left(\frac{h}{h_0}\right)^3 (\partial_s \psi^\ast) (\partial_s^2 \psi) \right)\right) .
\label{currad-sw}  
\eeq
We can easily prove that $\frac{|J_s^{-}|}{|J_s^{+}|} = e^{-\beta \omega}$, where $\beta = \frac{2 \pi}{\kappa}$. 
The grey-body factor can be obtained also in this case, analogously to the cases previously discussed, and the considerations we made for the BEC case and the cases discussed in \cite{belmaster} hold true also for the present case, 
so we don't discuss them again.\\ 
Again, the scattering involving  also the mode $v$, in the present framework, is of a different nature with respect to the 
one occurring for the production process of the Hawking mode $u$, and is associated with the scattering of 
Hawking modes on the background geometry provided by the reduced equation with $\delta=0$. We expect that 
also in this case the result is model-dependent, i.e.~it should depend strongly on the particular profiles one chooses for the different background  fields involved. 
Furthermore, the metric involved in the model studied in~\cite{coutant-sub} and taken into account herein corresponds to the case $\rho (x)=c^2 (x)$
of~\cite{coutant-parentani-fluid}, and the corresponding metric (2D part) is 
\beq
{ds}^2 = c(x) \bigl(c(x)^2 {dt}^2 -(dx - v(x) dt)^2\bigr),
\eeq
so that the equation one obtains when dispersion is neglected is just~\cite{coutant-parentani-fluid}
\beq
[(\partial_t +\partial_x v)(\partial_t +v \partial_x) -\partial_x c^2 \partial_x]\phi=0
\eeq
which amounts to the reduced equation. We do not calculate the backscattering contribution to the 
grey-body factor, instead we limit ourselves to notice that for $c(x)=$ const., as in the case of 
the subluminal Corley model~\cite{belmaster}, one would obtain $\Gamma=1$ up to higher order corrections, and then $R=0$. 
A truncation of the spectrum is to be meant for $\omega>\omega_{max}$ also in this case, 
as well known. $\omega_{max}$ has the same meaning as e.g.~in the BEC case.

We cannot claim that the present approach is the solution to the problem at hand, 
as a more detailed analysis of the experimental situations and of numerical simulations would be needed. Furthermore, we stress that 
we have not tried to take into account the so-called subcritical case~\cite{euve,parentani-sub,coutant-sub}, which requires a further 
analysis to be dealt with.

\section{Conclusions} 

We have first taken into account the BEC case and, under the assumption of small (but never vanishing) healing length we have found 
a scheme allowing to determine in a fully analytical way both the thermality and the grey-body factor of analogous Hawking radiation 
from BEC analogous black holes. Our analysis confirms that the master equation introduced in~\cite{belmaster} is actually effective also 
in the present case. We have also proposed an improvement in the near horizon analysis of the superluminal case, which is 
different from previous proposals like e.g.~\cite{corley,unruh-s}. In the second part of the paper, we have taken into account the 
the problem of surface gravity waves.  Thermality is again found in a simple way, and a scheme for the calculation of the 
grey-body factor is provided. Albeit the present framework provides a very interesting analytical picture of the Hawking effect in analogue gravity, it is by no means exhaustive and further analysis is required to delve into e.g the very low frequency regime in the case of surface waves.

\section*{Acknowledgements}

{\color{black} We thank Germain Rousseaux for sending us a list of references for completing our bibliography on water experiments and for sending us Ref. \cite{rousseaux-cocurrent}.} A.V.~was partially supported by MIUR-PRIN contract 2017CC72MK\_003.

\appendix

\section{Coefficients for the BEC equations}
\label{bec-coeffs}

\subsection{Equations for $\phi$}

\subsubsection{Coefficients of the simplified equation}

The coefficients which appears in~\eqref{four},
\begin{equation*}
\bigl[\alpha_4 (x) \partial_x^4 + \alpha_2 (x) \partial_x^2+ \alpha_1 (x) \partial_x +\alpha_0 (x)\bigr] \zeta (x)=0,
\end{equation*}
where $\phi (x) = c(x) \sqrt{v(x)} \zeta (x)$, are
\begin{equation*}
    \alpha_4 = \frac{\hbar ^2}{4 m^2 c^2(x)},
\end{equation*}

\begin{equation*}
     \alpha_3 =  -\frac{\hbar ^2 c'(x)}{m^2 c^3(x)}-\frac{\hbar ^2 v'(x)}{2 m^2 c^2(x) v(x)} ,
\end{equation*}

\begin{equation*}
\begin{split}
		\alpha_2 &=  \frac{\hbar ^2 c''(x)}{m^2 c^3(x)}-\frac{3 \hbar ^2 c'(x)^2}{2 m^2 c^4(x)}-\frac{i \hbar  v(x) c'(x)}{m c^3(x)}
	+\frac{\hbar ^2 v''(x)}{4 m^2 c^2(x) v(x)}\\
	&\quad -\frac{3 \hbar ^2 v'(x)^2}{8 m^2 c^2(x) v^2(x)}+\frac{i \hbar  v'(x)}{m c^2(x)}+\frac{v^2(x)}{c^2(x)}-1 ,
\end{split}
\end{equation*}

\begin{equation*}
	\begin{split}
		\alpha_1 &= \frac{\hbar ^2 c^{(3)}(x)}{m^2 c^3(x)}-\frac{i \hbar  v(x) c''(x)}{m c^3(x)}+\frac{3 \hbar ^2 c'(x)^3}{m^2 c(x)^5}-\frac{i \hbar  c'(x) v'(x)}{m c^3(x)}
		+\frac{i \hbar  v(x) c'(x)^2}{m c^4(x)}-\frac{2 \omega  \hbar  c'(x)}{m c^3(x)}-\frac{2 c'(x)}{c(x)}-\frac{4 \hbar ^2 c'(x) c''(x)}{m^2 c^4(x)}\\
		&\quad +\frac{\hbar ^2 v^{(3)}(x)}{4 m^2 c^2(x) v(x)}	+\frac{3 \hbar ^2 v'(x)^3}{4 m^2 c^2(x) v^3(x)}-\frac{\hbar ^2 v'(x) v''(x)}{m^2 c^2(x) v^2(x)}
		+\frac{i \hbar  v''(x)}{m c^2(x)}+\frac{2 v(x) v'(x)}{c^2(x)}-\frac{2 i \omega  v(x)}{c^2(x)} ,
	\end{split}
\end{equation*}

\begin{equation*}
	\begin{split}
		\alpha_0 &= \frac{\hbar ^2 c^{(4)}(x)}{4 m^2 c^3(x)}-\frac{\hbar ^2 c''(x)^2}{2 m^2 c^4(x)}+\frac{i \hbar  c''(x) v'(x)}{2 m c^3(x)}-\frac{\omega  \hbar  c''(x)}{m c^3(x)}
		+\frac{v^2(x) c''(x)}{c^3(x)}-\frac{c''(x)}{c(x)}-\frac{\hbar ^2 v^{(3)}(x) c'(x)}{4 m^2 c^3(x) v(x)}+\frac{\hbar ^2 c'(x)^2 v''(x)}{4 m^2 c^4(x) v(x)}\\
		&\quad -\frac{3 \hbar ^2 c'(x) v'(x)^3}{4 m^2 c^3(x) v^3(x)}-\frac{3 \hbar ^2 c'(x)^2 v'(x)^2}{8 m^2 c^4(x) v^2(x)}+\frac{\hbar ^2 c'(x) v'(x) v''(x)}{m^2 c^3(x) v^2(x)}
		+\frac{i \hbar  c'(x) v''(x)}{2 m c^3(x)}-\frac{3 i \hbar  c'(x) v'(x)^2}{4 m c^3(x) v(x)}-\frac{3 i \hbar  c'(x)^2 v'(x)}{2 m c^4(x)}\\
		&\quad +\frac{3 i \hbar  v(x) c'(x)^3}{m c(x)^5}+\frac{\omega  \hbar  c'(x)^2}{m c^4(x)}+\frac{v(x) c'(x) v'(x)}{c^3(x)}-\frac{2 v^2(x) c'(x)^2}{c^4(x)}
		-\frac{\hbar ^2 c^{(3)}(x) c'(x)}{m^2 c^4(x)}+\frac{3 \hbar ^2 c'(x)^2 c''(x)}{2 m^2 c(x)^5} \\
		&\quad -\frac{2 i \hbar  v(x) c'(x) c''(x)}{m c^4(x)}
		+\frac{\hbar ^2 v^{(4)}(x)}{8 m^2 c^2(x) v(x)}-\frac{7 \hbar ^2 v''(x)^2}{16 m^2 c^2(x) v^2(x)}-\frac{63 \hbar ^2 v'(x)^4}{64 m^2 c^2(x) v^4(x)}
		-\frac{5 \hbar ^2 v^{(3)}(x) v'(x)}{8 m^2 c^2(x) v^2(x)} \\
		&\quad +\frac{31 \hbar ^2 v'(x)^2 v''(x)}{16 m^2 c^2(x) v^3(x)}-\frac{3 i \hbar  v'(x)^3}{4 m c^2(x) v^2(x)}
		+\frac{i \hbar  v'(x) v''(x)}{m c^2(x) v(x)}+\frac{v(x) v''(x)}{2 c^2(x)}-\frac{i \omega  v'(x)}{c^2(x)}+\frac{v'(x)^2}{4 c^2(x)}\\
		&\quad -\frac{\omega ^2}{c^2(x)}-\frac{v''(x)}{2 v(x)}+\frac{3 v'(x)^2}{4 v^2(x)} . 
	\end{split}
\end{equation*}

\subsubsection{Coefficients expanded in powers of the healing length}

The coefficients which appears in~\eqref{orso},
\begin{equation*}
\bigl[\bar{\xi}^2\partial_x^4 -\bigl( 
\beta_2 (x,\bar{\xi}) \partial_x^2+ \beta_1 (x,\bar{\xi}) \partial_x +\beta_0 (x,\bar{\xi})\bigr)\bigr] \zeta (x)=0,
\end{equation*}
are
\begin{equation*}
	\beta_2 = \frac{2c^2(x)}{\bar{c}^2} \biggl[ \frac{v^2(x)}{c^2(x)}-1 +
        \left( -\frac{i \sqrt{2} \bar{c}  v(x) c'(x)}{c^3(x)}+\frac{i \sqrt{2} \bar{c} v'(x)}{c^2(x)}\right) \bar{\xi}+
	\left(\frac{\bar{c}^2 v''(x)}{2 c^2(x) v(x)}-\frac{3 \bar{c}^2 v'(x)^2}{4 c^2(x) v^2(x)}+\frac{2 \bar{c}^2  c''(x)}{c^3(x)}
        -\frac{3 \bar{c}^2  c'(x)^2}{c^4(x)}\right) \bar{\xi}^2 \biggr],
\end{equation*}

\begin{equation*}
	\begin{split}
		\beta_1 &= \frac{2c^2(x)}{\bar{c}^2} \biggl[-\frac{2 i \omega  v(x)}{c^2(x)}+\frac{2 v(x) v'(x)}{c^2(x)}-\frac{2 c'(x)}{c(x)}\\
                & \quad +\biggl(-\frac{i \sqrt{2} \bar{c}  v(x) c''(x)}{c^3(x)}-\frac{i \sqrt{2} \bar{c}  c'(x) v'(x)}{c^3(x)}+\frac{i \sqrt{2} \bar{c}  v(x) c'(x)^2}{c^4(x)}-\frac{2 \sqrt{2} \bar{c}  \omega  c'(x)}{c^3(x)} ++\frac{i \sqrt{2} \bar{c} v''(x)}{c^2(x)}\biggr) \bar{\xi}\\
                & \quad +\biggl(
\frac{\bar{c}^2  v^{(3)}(x)}{2 c^2(x) v(x)}+\frac{3 \bar{c}^2  v'(x)^3}{2 c^2(x) v^3(x)}-\frac{2 \bar{c}^2  v'(x) v''(x)}{c^2(x) v^2(x)}
		+\frac{2 \bar{c}^2  c^{(3)}(x)}{c^3(x)}
		-\frac{8 \bar{c}^2  c'(x) c''(x)}{c^4(x)} 
		 +\frac{6 \bar{c}^2  c'(x)^3}{c(x)^5}\biggr) \bar{\xi}^2 \biggr] ,
	\end{split}
\end{equation*}

\begin{equation*}
	\begin{split}
		\beta_0 &= \frac{2c^2(x)}{\bar{c}^2} \biggl[ \frac{v(x) v''(x)}{2 c^2(x)}-\frac{i \omega  v'(x)}{c^2(x)}+\frac{v'(x)^2}{4 c^2(x)}
		-\frac{\omega ^2}{c^2(x)}-\frac{v''(x)}{2 v(x)}+\frac{3 v'(x)^2}{4 v^2(x)}+\frac{v(x) c'(x) v'(x)}{c^3(x)}
		-\frac{2 v^2(x) c'(x)^2}{c^4(x)}\\
                & \quad +\frac{v^2(x) c''(x)}{c^3(x)}-\frac{c''(x)}{c(x)}
+\biggl(\frac{i \bar{c}  c''(x) v'(x)}{\sqrt{2} c^3(x)}-\frac{\sqrt{2} \bar{c}  \omega  c''(x)}{c^3(x)}	
		+\frac{i \bar{c}  c'(x) v''(x)}{\sqrt{2} c^3(x)}	-\frac{3 i \bar{c}  c'(x) v'(x)^2}{2 \sqrt{2} c^3(x) v(x)}-\frac{3 i \bar{c}  c'(x)^2 v'(x)}{\sqrt{2} c^4(x)}\\
		& \quad +\frac{3 i \sqrt{2} \bar{c} v(x) c'(x)^3}{c(x)^5}+\frac{\sqrt{2} \bar{c}  \omega  c'(x)^2}{c^4(x)}-\frac{2 i \sqrt{2} \bar{c}  v(x) c'(x) c''(x)}{c^4(x)}-\frac{3 i \bar{c}  v'(x)^3}{2 \sqrt{2} c^2(x) v^2(x)}+\frac{i \sqrt{2} \bar{c}  v'(x) v''(x)}{c^2(x) v(x)}\biggr) \bar{\xi}\\
& \quad +\biggl(
\frac{\bar{c}^2  v^{(4)}(x)}{4 c^2(x) v(x)}-\frac{7 \bar{c}^2  v''(x)^2}{8 c^2(x) v^2(x)}
		-\frac{63 \bar{c}^2  v'(x)^4}{32 c^2(x) v^4(x)}
		-\frac{5 \bar{c}^2 v^{(3)}(x) v'(x)}{4 c^2(x) v^2(x)}	+\frac{31 \bar{c}^2  v'(x)^2 v''(x)}{8 c^2(x) v^3(x)}-\frac{\bar{c}^2  c''(x)^2}{c^4(x)}\\
		& \quad 
		+\frac{\bar{c}^2  c^{(4)}(x)}{2 c^3(x)}-\frac{\bar{c}^2 v^{(3)}(x) c'(x)}{2 c^3(x) v(x)}+\frac{\bar{c}^2  c'(x)^2 v''(x)}{2 c^4(x) v(x)}	-\frac{3 \bar{c}^2  c'(x) v'(x)^3}{2 c^3(x) v^3(x)}-\frac{3 \bar{c}^2  c'(x)^2 v'(x)^2}{4 c^4(x) v^2(x)}
		+\frac{2 \bar{c}^2  c'(x) v'(x) v''(x)}{c^3(x) v^2(x)}\\
		& \quad +\frac{3 \bar{c}^2  c'(x)^2 c''(x)}{c(x)^5} 
		-\frac{2 \bar{c}^2  c^{(3)}(x) c'(x)}{c^4(x)}\biggr) \bar{\xi}^2 \biggr].
	\end{split}
\end{equation*}

\subsection{Equations for $\varphi$}

\subsubsection{Coefficients of the simplified equation}

The coefficients which appear in the equation
\begin{equation*}
\bigl[\gamma_4 (x) \partial_x^4 + \gamma_2 (x) \partial_x^2+ \gamma_1 (x) \partial_x +\gamma_0 (x)\bigr] \eta (x)=0,
\end{equation*}
where $\varphi (x) = c(x) \sqrt{v(x)} \eta (x)$, are
\begin{equation*}
    \gamma_4 = \frac{\hbar ^2}{4 m^2 c^2(x)},
\end{equation*}

\begin{equation*}
	\gamma_2 = \frac{\hbar ^2 c''(x)}{m^2 c^3(x)}-\frac{3 \hbar ^2 c'(x)^2}{2 m^2 c^4(x)}+\frac{i \hbar  v(x) c'(x)}{m c^3(x)}+\frac{\hbar ^2 v''(x)}{4 m^2 c^2(x) v(x)}
	-\frac{3 \hbar ^2 v'(x)^2}{8 m^2 c^2(x) v^2(x)}-\frac{i \hbar  v'(x)}{m c^2(x)}+\frac{v^2(x)}{c^2(x)}-1 ,
\end{equation*}

\begin{equation*}
	\begin{split}
		\gamma_1 &= \frac{\hbar ^2 c^{(3)}(x)}{m^2 c^3(x)}+\frac{i \hbar  v(x) c''(x)}{m c^3(x)}+\frac{3 \hbar ^2 c'(x)^3}{m^2 c(x)^5}+\frac{i \hbar  c'(x) v'(x)}{m c^3(x)}
		-\frac{i \hbar  v(x) c'(x)^2}{m c^4(x)}+\frac{2 \omega  \hbar  c'(x)}{m c^3(x)}-\frac{2 c'(x)}{c(x)} \\
		&\quad -\frac{4 \hbar ^2 c'(x) c''(x)}{m^2 c^4(x)}
		+\frac{\hbar ^2 v^{(3)}(x)}{4 m^2 c^2(x) v(x)}+\frac{3 \hbar ^2 v'(x)^3}{4 m^2 c^2(x) v^3(x)}-\frac{\hbar ^2 v'(x) v''(x)}{m^2 c^2(x) v^2(x)}
		-\frac{i \hbar  v''(x)}{m c^2(x)}+\frac{2 v(x) v'(x)}{c^2(x)}-\frac{2 i \omega  v(x)}{c^2(x)} ,
	\end{split}
\end{equation*}

\begin{equation*}
	\begin{split}
		\gamma_0 &= \frac{\hbar ^2 c^{(4)}(x)}{4 m^2 c^3(x)}-\frac{\hbar ^2 c''(x)^2}{2 m^2 c^4(x)}-\frac{i \hbar  c''(x) v'(x)}{2 m c^3(x)}+\frac{\omega  \hbar  c''(x)}{m c^3(x)}
		+\frac{v^2(x) c''(x)}{c^3(x)}-\frac{c''(x)}{c(x)}-\frac{\hbar ^2 v^{(3)}(x) c'(x)}{4 m^2 c^3(x) v(x)}+\frac{\hbar ^2 c'(x)^2 v''(x)}{4 m^2 c^4(x) v(x)}\\
		&\quad -\frac{3 \hbar ^2 c'(x) v'(x)^3}{4 m^2 c^3(x) v^3(x)}-\frac{3 \hbar ^2 c'(x)^2 v'(x)^2}{8 m^2 c^4(x) v^2(x)}+\frac{\hbar ^2 c'(x) v'(x) v''(x)}{m^2 c^3(x) v^2(x)}
		-\frac{i \hbar  c'(x) v''(x)}{2 m c^3(x)}+\frac{3 i \hbar  c'(x) v'(x)^2}{4 m c^3(x) v(x)}+\frac{3 i \hbar  c'(x)^2 v'(x)}{2 m c^4(x)}\\
		&\quad -\frac{3 i \hbar  v(x) c'(x)^3}{m c(x)^5}-\frac{\omega  \hbar  c'(x)^2}{m c^4(x)}+\frac{v(x) c'(x) v'(x)}{c^3(x)}-\frac{2 v^2(x) c'(x)^2}{c^4(x)}
		 -\frac{\hbar ^2 c^{(3)}(x) c'(x)}{m^2 c^4(x)}+\frac{3 \hbar ^2 c'(x)^2 c''(x)}{2 m^2 c(x)^5} \\
		 & \quad +\frac{2 i \hbar  v(x) c'(x) c''(x)}{m c^4(x)}
		\quad +\frac{\hbar ^2 v^{(4)}(x)}{8 m^2 c^2(x) v(x)}-\frac{7 \hbar ^2 v''(x)^2}{16 m^2 c^2(x) v^2(x)}-\frac{63 \hbar ^2 v'(x)^4}{64 m^2 c^2(x) v^4(x)}
		-\frac{5 \hbar ^2 v^{(3)}(x) v'(x)}{8 m^2 c^2(x) v^2(x)}\\
		& \quad +\frac{31 \hbar ^2 v'(x)^2 v''(x)}{16 m^2 c^2(x) v^3(x)}
		+\frac{3 i \hbar  v'(x)^3}{4 m c^2(x) v^2(x)}
		-\frac{i \hbar  v'(x) v''(x)}{m c^2(x) v(x)}+\frac{v(x) v''(x)}{2 c^2(x)}-\frac{i \omega  v'(x)}{c^2(x)}+\frac{v'(x)^2}{4 c^2(x)}
		-\frac{\omega ^2}{c^2(x)}-\frac{v''(x)}{2 v(x)}+\frac{3 v'(x)^2}{4 v^2(x)} .
	\end{split}
\end{equation*}

\subsubsection{Coefficients of the simplified equation with healing length}

The coefficients which appear in the equation
\begin{equation*}
\bigl[\bar{\xi}^2 \partial_x^4 - \bigl( \delta_2 (x,\bar{\xi}) \partial_x^2+ \delta_1 (x,\bar{\xi}) \partial_x +\delta_0 (x,\bar{\xi}) \bigr)\bigr] \eta (x)=0,
\end{equation*}
are

\begin{equation*}
	\begin{split}
	\delta_2 = \frac{2c^2(x)}{\bar{c}^2} \biggl[ \frac{v^2(x)}{c^2(x)}-1
	+\biggl(\frac{i \sqrt{2} \bar{c} v(x) c'(x)}{c^3(x)} -\frac{i \sqrt{2} \bar{c} v'(x)}{c^2(x)}\biggr) \bar{\xi}
	+\biggl(\frac{\bar{c}^2 v''(x)}{2 c^2(x) v(x)}-\frac{3 \bar{c}^2 v'(x)^2}{4 c^2(x) v^2(x)}+\frac{2 \bar{c}^2 c''(x)}{c^3(x)}
	-\frac{3 \bar{c}^2 c'(x)^2}{c^4(x)} \biggr) \bar{\xi}^2 \biggr],
\end{split}
\end{equation*}

\begin{equation*}
	\begin{split}
		\delta_1 &= \frac{2c^2(x)}{\bar{c}^2} \biggl[ 
		-\frac{2 c'(x)}{c(x)} +\frac{2 v(x) v'(x)}{c^2(x)}-\frac{2 i \omega  v(x)}{c^2(x)} \\
		&\quad +\biggl( \frac{i \sqrt{2} \bar{c} v(x) c''(x)}{c^3(x)} +\frac{i \sqrt{2} \bar{c} c'(x) v'(x)}{c^3(x)}
		-\frac{i \sqrt{2} \bar{c} v(x) c'(x)^2}{c^4(x)} +\frac{2 \sqrt{2} \bar{c} \omega  c'(x)}{c^3(x)}
		-\frac{i \sqrt{2} \bar{c} v''(x)}{c^2(x)} \biggr) \bar{\xi} \\		
		&\quad +\biggl( \frac{\bar{c}^2 v^{(3)}(x)}{2 c^2(x) v(x)}
		+\frac{3 \bar{c}^2 v'(x)^3}{2 c^2(x) v^3(x)}
		-\frac{2 \bar{c}^2 v'(x) v''(x)}{c^2(x) v^2(x)}
		+\frac{2 \bar{c}^2 \bar{c}^{(3)}(x)}{c^3(x)}
		-\frac{8 \bar{c}^2 c'(x) c''(x)}{c^4(x)}
		+\frac{6 \bar{c}^2 c'(x)^3}{c(x)^5} \biggr) \bar{\xi}^2
		 \biggr],
	\end{split}
\end{equation*}

\begin{equation*}
	\begin{split}
		\delta_0 &= \frac{2c^2(x)}{\bar{c}^2} \biggl[
		\frac{v^2(x) c''(x)}{c^3(x)} - \frac{c''(x)}{c(x)} + \frac{v(x) c'(x) v'(x)}{c^3(x)} - \frac{2 v^2(x) c'(x)^2}{c^4(x)}
		+ \frac{v(x) v''(x)}{2 c^2(x)} \\
		&\quad - \frac{i \omega  v'(x)}{c^2(x)} + \frac{v'(x)^2}{4 c^2(x)} - \frac{\omega ^2}{c^2(x)} - \frac{v''(x)}{2 v(x)} + \frac{3 v'(x)^2}{4 v^2(x)} \\
		&\quad + \biggl(-\frac{i \bar{c} c''(x) v'(x)}{\sqrt{2} c^3(x)}
		+\frac{\sqrt{2} \bar{c} \omega  c''(x)}{c^3(x)}
		-\frac{i \bar{c} c'(x) v''(x)}{\sqrt{2} c^3(x)}	
		+\frac{3 i \bar{c} c'(x) v'(x)^2}{2 \sqrt{2} c^3(x) v(x)}
		+\frac{3 i \bar{c} c'(x)^2 v'(x)}{\sqrt{2} c^4(x)} \\
		&\quad -\frac{3 i \sqrt{2} \bar{c} v(x) c'(x)^3}{c(x)^5}
		-\frac{\sqrt{2} \bar{c} \omega  c'(x)^2}{c^4(x)}
		+\frac{2 i \sqrt{2} \bar{c} v(x) c'(x) c''(x)}{c^4(x)}
		+\frac{3 i \bar{c} v'(x)^3}{2 \sqrt{2} c^2(x) v^2(x)}
		-\frac{i \sqrt{2} \bar{c} v'(x) v''(x)}{c^2(x) v(x)} \biggr) \bar{\xi} \\
		&\quad +\biggl(\frac{\bar{c}^2 v^{(4)}(x)}{4 c^2(x) v(x)}
		-\frac{7 \bar{c}^2 v''(x)^2}{8 c^2(x) v^2(x)}
		-\frac{63 \bar{c}^2 v'(x)^4}{32 c^2(x) v^4(x)}
		-\frac{5 \bar{c}^2 v^{(3)}(x) v'(x)}{4 c^2(x) v^2(x)}	
		+\frac{31 \bar{c}^2 v'(x)^2 v''(x)}{8 c^2(x) v^3(x)}
		-\frac{\bar{c}^2 c''(x)^2}{c^4(x)}	\\
		&\quad +\frac{\bar{c}^2 \bar{c}^{(4)}(x)}{2 c^3(x)}
		-\frac{\bar{c}^2 v^{(3)}(x) c'(x)}{2 c^3(x) v(x)}
		+\frac{\bar{c}^2 c'(x)^2 v''(x)}{2 c^4(x) v(x)}
		-\frac{3 \bar{c}^2 c'(x) v'(x)^3}{2 c^3(x) v^3(x)}
		-\frac{3 \bar{c}^2 c'(x)^2 v'(x)^2}{4 c^4(x) v^2(x)}
		+\frac{2 \bar{c}^2 c'(x) v'(x) v''(x)}{c^3(x) v^2(x)} \\
		&\quad +\frac{3 \bar{c}^2 c'(x)^2 c''(x)}{c(x)^5}
		-\frac{2 \bar{c}^2 \bar{c}^{(3)}(x) c'(x)}{c^4(x)} \biggr) \bar{\xi}^2  \biggr] .
	\end{split}
\end{equation*}

%%%%%%%%%%%%%%%%%%%


\begin{thebibliography}{99}

\bibitem{belmaster}
F.Belgiorno, S.L.Cacciatori and A.Vigan\`o, Analogue Hawking effect: a master equation. Preprint (2019). 

\bibitem{ni}

T.Nishimoto, K\u{o}dai Math. Sem. Rep. 29, (1978), 233.

\bibitem{ni-I}

T.Nishimoto, K\u{o}dai Math. Sem. Rep. 24, (1972), 281.

\bibitem{ni-tp}

T.Nishimoto, K\u{o}dai Math. Sem. Rep. 20, (1968), 218.

\bibitem{ni-global}

T.Nishimoto, K\u{o}dai Math. Sem. Rep. 27, (1976), 128.

\bibitem{rousseaux-first} 
  G.~Rousseaux, C.~Mathis, P.~Maissa, T.~G.~Philbin and U.~Leonhardt,
  %``Observation of negative phase velocity waves in a water tank: A classical analogue to the Hawking effect?,''
  New J.\ Phys.\  {\bf 10}, 053015 (2008)
  doi:10.1088/1367-2630/10/5/053015
  [arXiv:0711.4767 [gr-qc]].
  %%CITATION = doi:10.1088/1367-2630/10/5/053015;%%
  %133 citations counted in INSPIRE as of 12 May 2020

\bibitem{weinfurtner-prl} 
  S.~Weinfurtner, E.~W.~Tedford, M.~C.~J.~Penrice, W.~G.~Unruh and G.~A.~Lawrence,
  %``Measurement of stimulated Hawking emission in an analogue system,''
  Phys.\ Rev.\ Lett.\  {\bf 106}, 021302 (2011)
  doi:10.1103/PhysRevLett.106.021302
  [arXiv:1008.1911 [gr-qc]].

\bibitem{rousseaux-book} 
  J.~Chaline, G.~Jannes, P.~Maissa and G.~Rousseaux,
  %``Some aspects of dispersive horizons: lessons from surface waves,''
  Lect.\ Notes Phys.\  {\bf 870}, 145 (2013)
 % doi:10.1007/978-3-319-00266-8_7
  [arXiv:1203.2492 [physics.flu-dyn]].
  %%CITATION = doi:10.1007/978-3-319-00266-8_7;%%
  %4 citations counted in INSPIRE as of 24 May 2020



\bibitem{weinfurtner-book}
  S.~Weinfurtner, E.~W.~Tedford, M.~C.~J.~Penrice, W.~G.~Unruh and G.~A.~Lawrence,
  %``Classical aspects of Hawking radiation verified in analogue gravity experiment,''
  Lect.\ Notes Phys.\  {\bf 870}, 167 (2013)
  [arXiv:1302.0375 [gr-qc]].



\bibitem{rousseaux} 
  L.-P.~Euv\'e, F.~Michel, R.~Parentani, T.~G.~Philbin and G.~Rousseaux,
  %``Observation of noise correlated by the Hawking effect in a water tank,''
  Phys.\ Rev.\ Lett.\  {\bf 117}, no. 12, 121301 (2016)
  doi:10.1103/PhysRevLett.117.121301
  [arXiv:1511.08145 [physics.flu-dyn]].




\bibitem{jeff-nature} 
  J.~Steinhauer,
  %``Observation of self-amplifying Hawking radiation in an analog black hole laser,''
  Nature Phys.\  {\bf 10}, 864 (2014)
  doi:10.1038/NPHYS3104
  [arXiv:1409.6550 [cond-mat.quant-gas]].

\bibitem{denova} 
  J.~R.~Mu\~{n}oz de Nova, K.~Golubkov, V.~I.~Kolobov and J.~Steinhauer,
  %``Observation of thermal Hawking radiation and its temperature in an analogue black hole,''
  Nature {\bf 569}, no. 7758, 688 (2019)
  doi:10.1038/s41586-019-1241-0
  [arXiv:1809.00913 [gr-qc]].

\bibitem{rousseaux-cocurrent} 
  L.~P.~Euv\`e, S.~Robertson, N.~James, A.~Fabbri and G.~Rousseaux,
  %``Scattering of co-current surface waves on an analogue black hole,''
  Phys.\ Rev.\ Lett.\  {\bf 124}, no. 14, 141101 (2020)
  doi:10.1103/PhysRevLett.124.141101
  [arXiv:1806.05539 [gr-qc]].



\bibitem{corley}
S. Corley, % “Computing the spectrum of black hole radiation in the presence of high frequency dispersion: An analytical approach”, 
Phys. Rev. D, 57, 6280 (1998). 
[hep-th/9710075]. 




\bibitem{garay} 
  L.~J.~Garay, J.~R.~Anglin, J.~I.~Cirac and P.~Zoller,
  %``Black holes in Bose-Einstein condensates,''
  Phys.\ Rev.\ Lett.\  {\bf 85}, 4643 (2000)
  doi:10.1103/PhysRevLett.85.4643
  [gr-qc/0002015].

\bibitem{balbinot} 
  R.~Balbinot, A.~Fabbri, S.~Fagnocchi, A.~Recati and I.~Carusotto,
  %``Non-local density correlations as signal of Hawking radiation in BEC acoustic black holes,''
  Phys.\ Rev.\ A {\bf 78}, 021603 (2008)
  doi:10.1103/PhysRevA.78.021603
  [arXiv:0711.4520 [cond-mat.other]].

\bibitem{macher-bec} 
  J.~Macher and R.~Parentani,
  %``Black hole radiation in Bose-Einstein condensates,''
  Phys.\ Rev.\ A {\bf 80}, 043601 (2009)
  doi:10.1103/PhysRevA.80.043601
  [arXiv:0905.3634 [cond-mat.quant-gas]].

\bibitem{larre} 
  P.-E.~Larre, A.~Recati, I.~Carusotto and N.~Pavloff,
  %``Quantum fluctuations around black hole horizons in Bose-Einstein condensates,''
  Phys.\ Rev.\ A {\bf 85}, 013621 (2012)
  doi:10.1103/PhysRevA.85.013621
  [arXiv:1110.4464 [cond-mat.quant-gas]].

\bibitem{fabbri-hydro} 
  A.~Fabbri and C.~Mayoral,
  %``Step-like discontinuities in Bose-Einstein condensates and Hawking radiation: the hydrodynamic limit,''
  Phys.\ Rev.\ D {\bf 83}, 124016 (2011)
  doi:10.1103/PhysRevD.83.124016
  [arXiv:1004.4876 [gr-qc]].

\bibitem{mayoral-disp} 
  C.~Mayoral, A.~Fabbri and M.~Rinaldi,
  %``Step-like discontinuities in Bose-Einstein condensates and Hawking radiation: dispersion effects,''
  Phys.\ Rev.\ D {\bf 83}, 124047 (2011)
  doi:10.1103/PhysRevD.83.124047
  [arXiv:1008.2125 [gr-qc]].

\bibitem{anderson} 
  P.~R.~Anderson, R.~Balbinot, A.~Fabbri and R.~Parentani,
  %``Hawking radiation correlations in Bose Einstein condensates using quantum field theory in curved space,''
  Phys.\ Rev.\ D {\bf 87}, no. 12, 124018 (2013)
  doi:10.1103/PhysRevD.87.124018
  [arXiv:1301.2081 [gr-qc]].

\bibitem{balbinot-book} 
  R.~Balbinot, I.~Carusotto, A.~Fabbri, C.~Mayoral and A.~Recati,
  %``Understanding Hawking radiation from simple models of atomic Bose-Einstein condensates,''
  Lect.\ Notes Phys.\  {\bf 870}, 181 (2013)
  [arXiv:1207.2660 [gr-qc]].

\bibitem{anderson-gray} 
  P.~R.~Anderson, R.~Balbinot, A.~Fabbri and R.~Parentani,
  %``Gray-body factor and infrared divergences in 1D BEC acoustic black holes,''
  Phys.\ Rev.\ D {\bf 90}, no. 10, 104044 (2014)
  doi:10.1103/PhysRevD.90.104044
  [arXiv:1404.3224 [gr-qc]].

\bibitem{anderson-low} 
  P.~R.~Anderson, A.~Fabbri and R.~Balbinot,
  %``Low frequency gray-body factors and infrared divergences: rigorous results,''
  Phys.\ Rev.\ D {\bf 91}, no. 6, 064061 (2015)
  doi:10.1103/PhysRevD.91.064061
  [arXiv:1501.01953 [gr-qc]].

\bibitem{anderson-exact} 
  A.~Fabbri, R.~Balbinot and P.~R.~Anderson,
  %``Scattering coefficients and gray-body factor for 1D BEC acoustic black holes: exact results,''
  Phys.\ Rev.\ D {\bf 93}, no. 6, 064046 (2016)
  doi:10.1103/PhysRevD.93.064046
  [arXiv:1512.08447 [gr-qc]].


\bibitem{coutant-bdg} 
  A.~Coutant and S.~Weinfurtner,
  %``Low frequency analogue Hawking radiation: The Bogoliubov-de Gennes model,''
  Phys.\ Rev.\ D {\bf 97}, no. 2, 025006 (2018)
  doi:10.1103/PhysRevD.97.025006
  [arXiv:1707.09664 [gr-qc]].



\bibitem{eastham}

M.S.P. Eastham, The Asymptotic Solution of Linear Differential Systems: Application of the Levinson Theorem. 
London Mathematical Society Monographs New Series, Vol. 4. Clarendon Press, 1989.



\bibitem{unruh-s}

W.G. Unruh, and  R. Schutzhold, %“Universality of the Hawking effect”, 
Phys. Rev. D 71,
024028  (2005).  [gr-qc/0408009].

\bibitem{cpf} 
  A.~Coutant, R.~Parentani and S.~Finazzi,
  %``Black hole radiation with short distance dispersion, an analytical S-matrix approach,''
  Phys.\ Rev.\ D {\bf 85}, 024021 (2012)
  [arXiv:1108.1821 [hep-th]].

\bibitem{Coutant-thick} 
  
A.~Coutant and R.~Parentani,
  %``Hawking radiation with dispersion: The broadened horizon paradigm,''
  Phys.\ Rev.\ D {\bf 90}, no. 12, 121501 (2014)
  [arXiv:1402.2514 [gr-qc]].
 


\bibitem{miller}
P.D.Miller, {\it Applied Asymptotic Analysis}. Graduate Studies in Mathematics, Volume 75. American Mathematical Society, 
Providence, Rhode Island (2006).

\bibitem{dalfovo}
F.Dalfovo, A.Fracchetti, A.Lastri, L.Pitaevskii, and S.Stringari,
J. Low Temp. Phys. {\bf 104}, 367 (1996).




\bibitem{schutzhold} 
  R.~Schutzhold and W.~G.~Unruh,
%``Gravity wave analogs of black holes,''
  Phys.\ Rev.\ D {\bf 66}, 044019 (2002)
  doi:10.1103/PhysRevD.66.044019
  [gr-qc/0205099].



\bibitem{coutant-parentani-fluid} 
  A.~Coutant and R.~Parentani,
  %``Undulations from amplified low frequency surface waves,''
   Phys.\ Fluids {\bf 26}, 044106 (2014)
  arXiv:1211.2001 [physics.flu-dyn].

\bibitem{euve} 
  L.~P.~Euv\'e, F.~Michel, R.~Parentani and G.~Rousseaux,
  %``Wave blocking and partial transmission in subcritical flows over an obstacle,''
  Phys.\ Rev.\ D {\bf 91}, no. 2, 024020 (2015)
  doi:10.1103/PhysRevD.91.024020
  [arXiv:1409.3830 [gr-qc]].

\bibitem{parentani-sub} 
  S.~Robertson, F.~Michel and R.~Parentani,
  %``Scattering of gravity waves in subcritical flows over an obstacle,''
  Phys.\ Rev.\ D {\bf 93}, no. 12, 124060 (2016)
  doi:10.1103/PhysRevD.93.124060
  [arXiv:1604.07253 [gr-qc]].


\bibitem{coutant-sub} 
  A.~Coutant and S.~Weinfurtner,
  %``The imprint of the analogue Hawking effect in subcritical flows,''
  Phys.\ Rev.\ D {\bf 94}, no. 6, 064026 (2016)
  doi:10.1103/PhysRevD.94.064026
  [arXiv:1603.02746 [gr-qc]].




\bibitem{coutant-kdv} 
  A.~Coutant and S.~Weinfurtner,
  %``Low-frequency analogue Hawking radiation: The Korteweg\D0de Vries model,''
  Phys.\ Rev.\ D {\bf 97}, no. 2, 025005 (2018)
  doi:10.1103/PhysRevD.97.025005
  [arXiv:1707.09651 [gr-qc]].

\bibitem{prain} 
  M.~Richartz, A.~Prain, S.~Liberati and S.~Weinfurtner,
  %``Rotating black holes in a draining bathtub: superradiant scattering of gravity waves,''
  Phys.\ Rev.\ D {\bf 91}, no. 12, 124018 (2015)
  doi:10.1103/PhysRevD.91.124018
  [arXiv:1411.1662 [gr-qc]].

\bibitem{johnson-book}
R. S. Johnson, A Modern Introduction to the Mathematical Theory of Water Waves.
Cambridge Texts in Applied Mathematics Vol. 19. Cambridge University Press, 
Cambridge (1997). 


\end{thebibliography}
\end{document}